\documentclass[%
 reprint,
superscriptaddress,
 amsmath,amssymb,
 aps,
pra,
]{revtex4-2}

\usepackage{amsmath}
\usepackage{amssymb}
\usepackage{amsthm}
\usepackage{physics}
\usepackage{graphicx}
\usepackage{dcolumn}
\usepackage{bm}
\usepackage{hyperref}

\theoremstyle{definition}

\usepackage{xcolor}

\begin{document}

\preprint{APS/123-QED}

\title{Enhancing noise robustness in device-independent conference key agreement with asymmetric parity-CHSH inequalities}

\author{Makoto Ishihara}
 \affiliation{%
 Department of Electronics and Electrical Engineering, Keio University, 3-14-1 Hiyoshi, Kohoku-ku, Yokohama 223-8522, Japan
}%
\author{Wojciech Roga}%
 \affiliation{%
 Department of Electronics and Electrical Engineering, Keio University, 3-14-1 Hiyoshi, Kohoku-ku, Yokohama 223-8522, Japan
}%
\author{Jonatan Bohr Brask}
\affiliation{Center for Macroscopic Quantum States (bigQ), Department of Physics, Technical University of Denmark, 2800 Kongens Lyngby, Denmark}
\author{Masahiro Takeoka}%
 \email{takeoka@elec.keio.ac.jp}
\affiliation{%
 Department of Electronics and Electrical Engineering, Keio University, 3-14-1 Hiyoshi, Kohoku-ku, Yokohama 223-8522, Japan
}%
\affiliation{
National Institute of Information and Communications Technology (NICT), Koganei, Tokyo 184-8795, Japan.
}

\date{\today}

\begin{abstract}
Conference key agreement allows multiple remote parties to establish a shared secret key with information-theoretical security. In device-independent conference key agreement, security can be guaranteed with minimal assumptions on the devices used, provided that a violation of a Bell inequality is observed. However, implementations are extremely challenging because high detection efficiency is required to observe loophole-free Bell violations. Here, we enhance the robustness of device-independent conference key agreement by introducing a new family of multipartite Bell inequalities called the asymmetric parity-Clauser-Horne-Shimony-Holt (CHSH) inequalities. We derive a tight analytical lower bound on the conditional von Neumann entropy of the outcomes of one of the parties in a protocol based on this inequality, including noisy preprocessing. Using this bound, we analyze robustness to detection inefficiencies as well as local and global depolarizing noise. We show that the combination of the asymmetric parity-CHSH inequality and noisy preprocessing can significantly improve the robustness to imperfections.
\end{abstract}

\maketitle


\section{\label{sec:introduction}INTRODUCTION}
Conference key agreement (CKA) enables the distribution of a shared secret key among $N \geq 2$ remote parties with information-theoretical security~\cite{Murta2020}. CKA is one of the main applications of future quantum networks, and various CKA protocols have been proposed~\cite{Wu2016, Epping2017, Zhang2018, Grasselli2018, Ottaviani2019, Grasselli2019, Zhao2020, Carrara2023}. In addition, several experimental demonstrations of CKA protocols have been reported~\cite{Proietti2021, Pickston2023, Yang2024}. However, the security of CKA protocols relies on perfect characterization of the devices used for implementation. This is difficult to achieve in practice, which compromises the security of standard CKA protocols. On the other hand, device-independent CKA (DI-CKA) does not require any assumptions on the inner workings of the devices~\cite{Ribeiro2018, Holz2019, Ribeiro2019, Holz2020, Grasselli2023, Ishihara2025, Ishihara2026}. Security can be established from observations of Bell-inequality violations which guarantee the existence of nonlocal correlations among the remote parties and limit the information that any eavesdropper can possess. 

While DI-CKA protocols enjoy ultimate security, experimental implementation of existing protocols is extremely challenging. The reason is that DI-CKA protocols require loophole-free Bell-inequality violations, and in particular violations without any postselection, which makes them very sensitive to detection efficiency and noise. Therefore, enhancing the robustness against experimental imperfections is necessary for the practical realization of DI-CKA protocols.

One way to achieve this is by using classical postprocessing methods which can lower the threshold detection efficiency~\cite{Tan2020, Ho2020, Xu2022}. Among such methods, noisy preprocessing is widely adopted in recent studies. In noisy preprocessing, one party, say Alice, flips her key bits with some probability~\cite{Ho2020}. This procedure reduces the information of an eavesdropper, Eve, about Alice's system and makes the protocol more robust against experimental imperfections.

Another way is to use more sophisticated Bell inequalities. In the case of device-independent quantum key distribution (DI-QKD), which is equivalent to DI-CKA with $N=2$, the most standard Bell inequality is the well-known Clauser-Horne-Shimony-Holt (CHSH) inequality~\cite{Clauser1969}. However, Woodhead et al.\ introduced the asymmetric CHSH inequality, a generalization of the CHSH inequality which accounts for asymmetry between measurements in DI-QKD and leads to an improvement in the robustness against experimental noise~\cite{Woodhead2021}. The parity-CHSH inequality developed in Ref.~\cite{Ribeiro2019} is a multipartite generalization of CHSH which can be used for DI-CKA. However, this inequality does not incorporate asymmetry and can be improved further.

In this paper, we propose a DI-CKA protocol which is robust against experimental imperfections. We achieve this by introducing a new multipartite Bell inequality which we call the asymmetric parity-CHSH inequality and combining it with noisy preprocessing. The asymmetric parity-CHSH inequality is a multipartite Bell inequality which generalizes the parity-CHSH inequality to incorporate the asymmetric behavior of measurements. Our main technical contribution is a tight lower bound on the conditional von Neumann entropy of outcomes of one of the parties (Alice) for the asymmetric parity-CHSH inequality with noisy preprocessing. Ribeiro et al.\ derived a lower bound on the conditional von Neumann entropy for the parity-CHSH inequality~\cite{Ribeiro2019}, and Grasselli et al. proved that the bound is tight~\cite{Grasselli2023}. However, they did not consider the asymmetry in the protocol nor noisy preprocessing, and thus the resulting bound was less robust to imperfections. To derive our lower bound, we define a nonlocal game which is equivalent to the asymmetric CHSH inequality and extends it to the multipartite case. By using the derived bound, we analyze the effects of experimental imperfections, in particular detection efficiency, global depolarizing noise, and local depolarizing noise, and we show that the combination of the asymmetric parity-CHSH inequality and noisy preprocessing can significantly improve the performance of the DI-CKA protocol. 

The rest of the paper is structured as follows. In Sec.~\ref{section:parity}, we introduce the asymmetric parity-CHSH inequality. Next, in Sec.~\ref{section:DICKAprotocol}, we describe the DI-CKA protocol. In Sec.~\ref{section:derivinglowerbound}, we derive the lower bound on the conditional von Neumann entropy of Alice's outcomes, and in Sec.~\ref{section:tightness}, we show that the bound is tight. In Sec.~\ref{section:imperfection} we analyze the robustness of our DI-CKA protocol against experimental imperfections and finally conclude in Sec.~\ref{section:conclusion}.

\section{ASYMMETRIC PARITY-CHSH INEQUALITY}\label{section:parity}
In this section we introduce the asymmetric parity-CHSH inequality. To do so, we first define a nonlocal game which is equivalent to the asymmetric CHSH inequality. Then, we generalize it to a multipartite nonlocal game which is equivalent to the asymmetric parity-CHSH inequality.

\subsection{Asymmetric CHSH game}
Assume that two parties, Alice and Bob, try to violate the CHSH inequality defined in the following. Let $A_x$ denote Alice's binary observable, where $x \in \{ 0, 1 \}$ denotes the measurement input and $a \in \{ 0, 1 \}$ is the corresponding measurement outcome. We associate each outcome $a$ with the value $(-1)^a$. Similarly, let $B_y$ be Bob's measurement with outcome $b \in \{ 0, 1 \}$ and measurement input $y \in \{ 0, 1 \}$. Then, the CHSH inequality~\cite{Clauser1969} is expressed as follows
\begin{equation}\label{eq:CHSHineq}
    S^\text{CHSH} = \ev{A_0 (B_0 + B_1)} + \ev{A_1 (B_0 - B_1)} \leq 2,
\end{equation}
where we call $S^\text{CHSH}$ a CHSH value. The classical limit of this value is 2, while the quantum limit is $2 \sqrt{2}$. 

Consider a nonlocal game called a CHSH game which is equivalent to the CHSH inequality as follows. Let a referee select a pair of questions $(x, y) \in \{ 0, 1\}$ and send $x$ to Alice and $y$ to Bob according to a probability distribution $\mathbf{p} (x,y)$ where
\begin{equation}
        \begin{split}
            &\mathbf{p}(0, 0) = \mathbf{p}(0, 1)=\mathbf{p}(1,0)=\mathbf{p}(1,1)
            =\frac{1}{4}.
        \end{split}
\end{equation}
Given the questions, Alice (Bob) answers a bit $a \in \{0, 1\}$ ($b \in \{0, 1\}$). Then, Alice and Bob win this game if and only if
\begin{equation}\label{eq:WinConditionCHSH}
        a +  b = xy \, \text{mod 2}.
\end{equation}
The winning probability of this game $P^{\text{CHSH}}_{\text{win}}$ with classical strategies satisfies the following inequality
\begin{equation}
        P^{\text{CHSH}}_{\text{win}} \leq \frac{3}{4}.
\end{equation}
When Alice and Bob use quantum strategies, the maximum winning probability becomes
\begin{equation}
    P^{\text{CHSH}}_{\text{win}} \leq \frac{2+\sqrt{2}}{4}.
\end{equation}
The winning probability of the CHSH game relates to the CHSH value $S^\text{CHSH}$ in the following way:
\begin{equation}\label{eq:CHSHrelation}
    P_\text{win}^\text{CHSH} = \frac{S^\text{CHSH}}{8} + \frac{1}{2}.
\end{equation}
We derive this relationship in Appendix~\ref{appendix:relationship}. 

The asymmetric CHSH value is as follows~\cite{Lawson2010, Acin2012,Woodhead2021}
\begin{equation}
    S^\text{CHSH}_\alpha = \alpha \ev{A_0 (B_0 + B_1)} + \ev{A_1 (B_0 - B_1)},
\end{equation}
where $\alpha > 0$. When $\alpha = 1$, this is equivalent to the CHSH value $S^\text{CHSH}$. The classical limit and the quantum limit of the asymmetric CHSH value $S^\text{CHSH}_\alpha$ are
\begin{equation}
    \begin{split}
        S^\text{CHSH}_\alpha &\leq \begin{cases}
            2\alpha & \text{if} \, \alpha \geq 1\\
            2 & \text{if} \, 0 < \alpha <1
        \end{cases}\\
        &\leq 2 \sqrt{1+\alpha^2}.
    \end{split}
\end{equation}

Let us define the corresponding asymmetric CHSH game. While it is similar to the CHSH game, in this version a referee selects a pair of questions with a biased probability distribution:
\begin{align}
            \mathbf{p}(0, 0) = \mathbf{p}(0, 1)&=\frac{\alpha}{2(1+\alpha)} = \frac{1}{4}\frac{2\alpha}{1+\alpha} ,\\
            \mathbf{p}(1, 0) = \mathbf{p}(1, 1)&=\frac{1}{2(1+\alpha)} = \frac{1}{4}\frac{2}{1+\alpha} .
\end{align}
Alice and Bob win this game under the same condition as in the CHSH game shown in Eq.~(\ref{eq:WinConditionCHSH}). The winning probability of this game $P^{\text{ACHSH}}_{\text{win}}$ with classical strategies satisfies the following inequality
\begin{equation}\label{eq:ClassicalWinProbACHSH}
    \begin{split}
        P^{\text{ACHSH}}_{\text{win}} \leq \begin{cases}
        \frac{1+2\alpha}{2(1+\alpha)}& \text{if} \, \alpha \geq 1,\\
        \frac{2+\alpha}{2(1+\alpha)} & \text{if} \, 0<\alpha < 1.
        \end{cases}
    \end{split}
\end{equation}
When the players employ quantum strategies, the maximal winning probability becomes
\begin{equation}\label{eq:QuantumWinProbACHSH}
        P^\text{ACHSH}_\text{win} \leq \frac{\sqrt{1+\alpha^2}}{2(1+\alpha)} + \frac{1}{2},
\end{equation}
and it is related to the asymmetric CHSH value $S_\alpha^\text{CHSH}$ as follows (see Appendix~\ref{appendix:relationship} for details):
\begin{equation}\label{eq:ACHSHrelation}
    P^\text{ACHSH}_\text{win} = \frac{S_\alpha^\text{CHSH}}{4 (1 + \alpha)} + \frac{1}{2}.
\end{equation}
We provide explicit strategies saturating the classical and quantum limits of $P_\text{win}^\text{ACHSH}$ in Appendix~\ref{appendix:ExplicitStrategies}.

\subsection{Asymmetric parity-CHSH inequality}
The parity-CHSH inequality is an extension of the CHSH inequality to more than two parties. We consider $N$ parties, Alice, $\text{Bob}_1, \ldots, \text{Bob}_{N-1}$, and Alice and $\text{Bob}_1$ each have binary measurement inputs and outputs. Each of the other $N-2$ parties has a single input and a binary output. Let $A_x$ and $B_{y_1}^1$ denote Alice's measurement with input $x \in \{0, 1 \}$ and output $a \in \{ 0, 1 \}$, and $\text{Bob}_1$'s measurement with input $y_1 \in \{ 0, 1 \}$ and output $b_1 \in \{ 0, 1 \}$, respectively. Let $B^i_{y_i}$ denote $\text{Bob}_i$'s measurement with input $y_i = 1$ and output $b_i \in \{ 0, 1\}$ for $2 \leq i \leq N-1$. Then, the parity-CHSH inequality is expressed in the following way:
\begin{equation}\label{eq:ParityCHSHinequality}
\begin{split}
    S^\text{P-CHSH} &= \ev{A_0 B_{+}^1} + \ev{A_1 B_{-}^1 \bigotimes_{i = 2}^{N-1} B^i_1} \leq 2,
\end{split}
\end{equation}
where $B_\pm^1 = B_0^1 \pm B_1^1$, and we call $S^\text{P-CHSH}$ a parity-CHSH value. The quantum limit of $S^\text{P-CHSH}$ is $2 \sqrt{2}$. The parity-CHSH inequality can also be expressed as a nonlocal game. We define a parity-CHSH game as follows. A referee selects a pair of questions $(x, y_1) \in \{0, 1 \}$ according to a probability distribution $\mathbf{p} (x, y_1)$ where
\begin{equation}
        \mathbf{p} (0, 0) = \mathbf{p} (0, 1) = \mathbf{p} (1, 0) = \mathbf{p} (1, 1) = \frac{1}{4}.
\end{equation}
The referee sends $x$ to Alice, $y_1$ to $\text{Bob}_1$, and a fixed question $y_i = 1$ to $\text{Bob}_i$ for $2 \leq i \leq N-1$. Given the questions, Alice answers a bit $a \in \{0, 1\}$ and $\text{Bob}_i$ answers a bit $b_i \in \{0, 1\}$ for $1 \leq i \leq N-1$. Here, let $\bar{b} = \oplus_{2 \leq i \leq N-1} b_i$ denote the parity of all the answers of $\text{Bob}_2, \ldots, \text{Bob}_{N-1}$. The $N$ players win this game if and only if
\begin{equation}\label{eq:WinConditionPCHSH}
        a + b_1 = x (y_1+\bar{b}) \, \text{mod} \, 2.
\end{equation}
When the players employ classical strategies, the winning probability of this parity-CHSH game $P_\text{win}^\text{P-CHSH}$ satisfies the following inequality
    \begin{equation}
        P_{\text{win}}^\text{P-CHSH} \leq \frac{3}{4},
    \end{equation}
and $P_\text{win}^\text{P-CHSH} \leq \frac{2+\sqrt{2}}{4}$ with quantum strategies. Importantly, when $\bar{b} = 0$, the parity-CHSH game is equivalent to the CHSH game. Also, when $\bar{b} = 1$, this game is equivalent to the CHSH game up to relabelling $\text{Bob}_1$'s input $y_1$. We can derive the following relationship between the winning probability of the parity-CHSH game $P^\text{P-CHSH}_\text{win}$ and the parity-CHSH value $S^\text{P-CHSH}$ same as that of the CHSH inequality (see Appendix~\ref{appendix:relationship}).
\begin{equation}\label{eq:Parityrelation}
    P_\text{win}^\text{P-CHSH} = \frac{S^\text{P-CHSH}}{8} + \frac{1}{2}.
\end{equation}

Finally, we define an asymmetric parity-CHSH value $S_\alpha^\text{P-CHSH}$ which is an extension of the parity-CHSH value as follows.
\begin{equation}\label{eq:AsymmetricParityCHSHinequality}
\begin{split}
    S_\alpha^\text{P-CHSH} &= \alpha \ev{A_0 B_{+}^1} + \ev{A_1 B_{-}^1 \bigotimes_{i = 2}^{N-1} B_1^i},
\end{split}
\end{equation}
This is equivalent to the parity-CHSH value when $\alpha = 1$, and its classical and quantum limits can be expressed as follows
\begin{equation}
    \begin{split}
        S^\text{P-CHSH}_\alpha &\leq \begin{cases}
            2\alpha & \text{if} \, \alpha \geq 1\\
            2 & \text{if} \, 0 < \alpha <1
        \end{cases}\\
        &\leq 2 \sqrt{1+\alpha^2},
    \end{split}
\end{equation}
which are the same as those of the asymmetric CHSH value $S^\text{CHSH}_\alpha$. 

We can define a nonlocal game which is equivalent to the asymmetric parity-CHSH inequality called an asymmetric parity-CHSH game. In this game, a referee selects a pair of questions $(x, y_1) \in \{0, 1\}$ according to the following biased probability distribution
\begin{align}
        \mathbf{p} (0, 0) = \mathbf{p} (0, 1) =  \frac{\alpha}{2(1+\alpha)},\\
        \mathbf{p} (1, 0) = \mathbf{p} (1, 1) =  \frac{1}{2(1+\alpha)}.
\end{align}
A winning condition of this game is the same as that of the parity-CHSH game shown in Eq.~(\ref{eq:WinConditionPCHSH}). The winning probability of this game with classical strategies $P_\text{win}^\text{AP-CHSH}$ satisfies
\begin{equation}
    \begin{split}
        P^{\text{AP-CHSH}}_{\text{win}} \leq \begin{cases}
        \frac{1+2\alpha}{2(1+\alpha)} & \text{if} \, \alpha \geq 1\\
        \frac{2+\alpha}{2(1+\alpha)} & \text{if} \, 0<\alpha < 1
        \end{cases}
    \end{split}
\end{equation}
The winning probability with quantum strategies satisfies the following
\begin{equation}
        P^\text{AP-CHSH}_\text{win} \leq \frac{\sqrt{1+\alpha^2}}{2(1+\alpha)} + \frac{1}{2}.
\end{equation}
The relationship between the winning probability of the asymmetric parity-CHSH game $P^\text{AP-CHSH}_\text{win}$ and the asymmetric parity-CHSH value $S^\text{P-CHSH}_\alpha$ can be written as
\begin{equation}\label{eq:APCHSHrelation}
    P_\text{win}^\text{AP-CHSH} = \frac{S_\alpha^\text{P-CHSH}}{4(1+\alpha)} + \frac{1}{2}.
\end{equation}
We provide a derivation of this relationship in Appendix~\ref{appendix:relationship}. Note that the asymmetric parity-CHSH game is equivalent to the asymmetric CHSH game when $\bar{b} = 0$. In addition, the asymmetric parity-CHSH game is equivalent to the asymmetric CHSH game up to relabelling $\text{Bob}_1$'s input $y_1$ when $\bar{b}=1$.

\section{DI-CKA PROTOCOL}\label{section:DICKAprotocol}
Let us describe a DI-CKA protocol which we consider throughout this paper. $N$ legitimate parties (Alice, $\text{Bob}_1, \ldots, \text{Bob}_{N-1}$) try to share a common secret key. First, an $N$-partite Greenberger-Horne-Zeilinger (GHZ) state is distributed among the parties. Then, each party performs some measurements on the distributed GHZ state. Let $A_x$ ($B_{y_1}^1$) be Alice's ($\text{Bob}_1$'s) measurement with a binary outcome $a \in \{ 0, 1\}$ ($b_1 \in \{0, 1\}$) for input $x \in \{0, 1\}$ ($y_1 \in \{ 0, 1, 2\}$). Let $B_{y_i}^i$ denote $\text{Bob}_i$'s measurement with a binary outcome $b_i \in \{0, 1 \}$ for input $y_i \in \{1, 2 \}$ for $2 \leq i \leq N-1$. A subset of rounds where Alice chooses $x = 0$ and $\text{Bob}_i$ chooses $y_i = 2$ for all $i$ is classified as key-generation rounds and used to generate a common secret key. The remaining rounds are classified as Bell-test rounds and used to calculate the asymmetric parity-CHSH value $S^\text{P-CHSH}_\alpha$ and a key rate. If the parameter estimation shows a positive key rate, the parties perform error correction and privacy amplification, resulting in a shared secret key.

In addition, the parties perform noisy preprocessing to enhance the robustness of the protocol against imperfections such as detection efficiency~\cite{Ho2020}. Here, Alice intentionally introduces noise to her raw key bits by flipping each bit with a probability $q$. This procedure reduces the correlation between Alice and any potential eavesdropper, Eve, since Eve does not know which bits are flipped. On the other hand, it also reduces the correlation between Alice and the other parties. For an appropriate choice of the probability $q$, the net effect can be positive, making the DI-CKA protocol more robust against imperfections. 

The asymptotic key rate of the DI-CKA protocol $K$ can be written in the following way~\cite{Ribeiro2018}
\begin{equation}\label{eq:keyrate}
    K = H(A|E, x^*) - \max_{1 \leq i \leq N-1} H(A|B_i, x^*, y^*_i),
\end{equation}
where $A, B_i$ and $E$ denote Alice's, $\text{Bob}_i$'s and Eve's systems, respectively, and $x^* = 0$ ($y_i^* = 2$) denotes the measurement input of Alice ($\text{Bob}_i$) for the key-generation rounds. The first term in Eq. (\ref{eq:keyrate}) is the conditional von Neumann entropy of Alice's measurement outcomes for the key-generation rounds given Eve's information. The second term is the conditional Shannon entropy between Alice and $\text{Bob}_i$ corresponding to the error correction cost. The conditional Shannon entropy can be calculated directly from the probability distribution $P(a, b_1, \ldots, b_{N-1}|x, y_1, \ldots, y_{N-1})$ for the Bell test rounds. On the other hand, the calculation of the first term is not straightforward since it contains Eve's uncharacterized system. Although it is possible to calculate this term by using numerical optimization methods~\cite{Masanes2011, Brown2021, Tan2021, Brown2024}, such methods require heavy computational costs. Therefore, in the following sections, we derive a tight analytical lower bound on the term which depends on the asymmetric parity-CHSH value $S^\text{P-CHSH}_\alpha$, the parameter corresponding to the asymmetry $\alpha$, and the probability of noisy preprocessing $q$.

\section{ANALYTICAL LOWER BOUND}\label{section:derivinglowerbound}
We derive an analytical lower bound on the conditional von Neumann entropy in Eq. (\ref{eq:keyrate}). From the properties of the conditional von Neumann entropy, the following holds
\begin{equation}
    \begin{split}
        H(A|E, x^*) &\geq H(A|E, x^*, \bar{b})\\
        &= P_{\bar{b}=0} H(A|E,x^*, \bar{b}=0) \\
        &+ P_{\bar{b} = 1} H(A|E, x^*, \bar{b}=1),
    \end{split}
\end{equation}
where $P_{\bar{b}=0}$ ($P_{\bar{b}=1}$) denotes the probability that $\bar{b}=0$ ($\bar{b}=1$). Since the asymmetric parity-CHSH game is equivalent to the asymmetric CHSH game when $\bar{b} = 0$, we have $P_{\text{win}|\bar{b}=0}^\text{AP-CHSH} = P_\text{win}^\text{A-CHSH}$, where $P_{\text{win}|\bar{b}=0}^\text{AP-CHSH}$ represents the winning probability of the asymmetric parity-CHSH game for $\bar{b} = 0$. Then, we can bound $H(A|E,x^*, \bar{b}=0)$ with the tight lower bound developed for the asymmetric CHSH inequality~\cite{Woodhead2021}:
\begin{equation}
    \begin{split}
        H(A|E,x^*, \bar{b}=0) \geq \bar{g}_{q, \alpha}(P_{\text{win}|\bar{b}=0}^\text{AP-CHSH}),
    \end{split}
\end{equation}
We describe the definition of the function $\bar{g}_{q, \alpha}$ later. We can analogously lower bound $H(A|E, x^*, \bar{b} = 1)$, obtaining 
\begin{equation}
    \begin{split}
        H(A|E,x^*) &\geq P_{\bar{b}=0} \bar{g}_{q, \alpha}(P^\text{AP-CHSH}_{\text{win}|\bar{b}=0}) \\
        &+ P_{\bar{b}=1} \bar{g}_{q, \alpha}(P^\text{AP-CHSH}_{\text{win}|\bar{b}=1}),
    \end{split}
\end{equation}
where $P_{\text{win}|\bar{b}=1}^\text{AP-CHSH}$ denotes the winning probability of the asymmetric parity-CHSH game when $\bar{b}=1$. By using the convexity of the function $\bar{g}_{q, \alpha}$~\cite{Woodhead2021}, we have the following
\begin{equation}
    \begin{split}
        H(A|E, x^*) 
        &\geq \bar{g}_{q, \alpha}(P_{\bar{b}=0} P^\text{AP-CHSH}_{\text{win}|\bar{b}=0} + P_{\bar{b}=1} P^\text{AP-CHSH}_{\text{win} |\bar{b}=1} )\\
        &= \bar{g}_{q, \alpha}(P^\text{AP-CHSH}_\text{win}).
    \end{split}
\end{equation}

Here, we describe the definition of the function $\bar{g}_{q, \alpha}$. Since there is the relationship between the winning probability of the asymmetric parity-CHSH game $P^\text{AP-CHSH}_\text{win}$ and the asymmetric parity-CHSH value $S^\text{P-CHSH}_\alpha$ shown in Eq. (\ref{eq:APCHSHrelation}), we can express the function in terms of a Bell value $S$. The function $\bar{g}_{q, \alpha}(S)$ is defined as follows
\begin{widetext}
\begin{equation}\label{eq:lowerboundBell}
    \bar{g}_{q, \alpha} (S) = \begin{cases}
        g_{q, \alpha}(S) &\text{if} \, \alpha \geq 1 \, \text{or} \, S \geq S_* \\
        h(q) + g'_{q, \alpha}(S_*) (|S|-2) &\text{if} \, 0<\alpha <1 \, \text{and} \, S < S_*
    \end{cases},
\end{equation}
where
\begin{equation}
    \begin{split}
        g_{q, \alpha}(S) = 1 - h\left( \frac{1 + \sqrt{S^2/4 - \alpha^2}}{2} \right) + h \left( \frac{1 + \sqrt{1 + 4q (1-q) (S^2/4 - \alpha^2 -1)}}{2} \right),
    \end{split}
\end{equation}
\end{widetext}
$h(x) = -x \log_2 x - (1-x)\log_2 (1-x)$ is the binary entropy and $g'_{q, \alpha}$ is the first derivative of $g_{q, \alpha}$. $S_*$ is the unique point satisfying the following
\begin{equation}
    h(q) + g'_{q, \alpha} (S_*) (S_* - 2) = g_{q, \alpha} (S_*),
\end{equation}
that is, the point where the tangent of $g_{q, \alpha} (S)$ crosses $h(q)$ at $S = 2$. Note that we can express the function $\bar{g}_{q, \alpha}$ in terms of the winning probability of the asymmetric parity-CHSH game by substituting the relationship in Eq.~(\ref{eq:APCHSHrelation}) into Eq.~(\ref{eq:lowerboundBell}).

\section{TIGHTNESS OF LOWER BOUND}\label{section:tightness}
In this section, we show that the lower bound we derive in the previous section is tight. To this end, we consider an explicit eavesdropping attack by Eve which is similar to the optimal eavesdropping attack for the asymmetric CHSH inequality proposed in Ref.~\cite{Woodhead2021}. First, we consider the following quantum state shared among the $N$ parties and Eve
\begin{equation}
    \rho_{A\boldsymbol{B}E} = \ketbra{\Psi}{\Psi}_{A\boldsymbol{B}E},
\end{equation}
where $\ket{\Psi}$ is a pure state of the form
\begin{equation}
    \ket{\Psi}_{A\boldsymbol{B}E} = \frac{1}{\sqrt{2}} (\ket{00\cdots 0}_{A\boldsymbol{B}} \ket{\psi_0}_E + \ket{11\cdots 1}_{A\boldsymbol{B}} \ket{\psi_1}_E),
\end{equation}
$\boldsymbol{B}$ denotes the Bobs' systems, $B_1, \ldots, B_{N-1}$, and $\ket{\psi_0}$ and $\ket{\psi_1}$ satisfy
\begin{equation}
    \braket{\psi_0}{\psi_1} = F \in [ 0, 1].
\end{equation}
We also assume that the parties perform the following measurements
\begin{align}
    A_0 &= \sigma_Z, \quad A_1 = \sigma_X,\\
    B_0^1 &= \cos \theta \sigma_Z + \sin \theta \sigma_X,\\
    B_1^1 &= \cos \theta \sigma_Z - \sin \theta \sigma_X,\\
    B^2_1 &= \cdots =B^{N-1}_1 = \sigma_X.
\end{align}

Using the definition of the conditional von Neumann entropy $H(A|E,x^*)$, we have
\begin{equation}
\label{eq:condHrewrite}
    \begin{split}
        H(A|E,x^*) = H(A|x^*) - H(E) + H(E|A,x^*).
    \end{split}
\end{equation}
Since Alice measures $\sigma_Z$ in the key-generation rounds ($x^* = 0)$, for the first term on the right-hand side, we have $H(A|x^*) = 1$. We now compute the second and third terms.

Since Eve's quantum state is
\begin{equation}
    \begin{split}
        \rho_E = \text{Tr}_{A\boldsymbol{B}} [\ketbra{\Psi}{\Psi}_{A\boldsymbol{B}E}] = \frac{1}{2} (\ketbra{\psi_0}{\psi_0} + \ketbra{\psi_1}{\psi_1}) ,
    \end{split}
\end{equation}
defining $\ket{\psi_0} = \ket{0}$ and $\ket{\psi_1} = F\ket{0} + \sqrt{1-F^2} \ket{1}$, the eigenvalues of $\rho_E$ become $(1 \pm F)/2$. Therefore, 
\begin{equation}
    H(E) = h\left(\frac{1+F}{2} \right).
\end{equation}

Considering Alice's $\sigma_Z$ measurement, the classical-quantum state between Alice's measurement outcomes and Eve's system $\rho_{aE}$ can be written as
\begin{equation}
    \begin{split}
        \rho_{aE} &= \sum_{a=0}^1 \ketbra{a}{a} \otimes \text{Tr}_{A \boldsymbol{B}} [\ketbra{a}{a}_A \ketbra{\Psi}{\Psi}_{A \boldsymbol{B}E}]\\
        &= \sum_{a=0}^1 \frac{1}{2} \ketbra{a}{a} \otimes \ketbra{\psi_a}{\psi_a}_E.
    \end{split}
\end{equation}
Let $\tilde{\rho}_{aE}$ denote the quantum state after performing noisy preprocessing on $\rho_{aE}$. Then,
\begin{equation}
\begin{split}
    \tilde{\rho}_{aE} &= \frac{1}{2} \ketbra{0}{0} \otimes ((1-q) \ketbra{\psi_0}{\psi_0} + q \ketbra{\psi_1}{\psi_1})\\
    &+ \frac{1}{2} \ketbra{1}{1} \otimes ((1-q)\ketbra{\psi_1}{\psi_1} + q \ketbra{\psi_0}{\psi_0})\\
    &= \sum_{a=0}^1 \frac{1}{2} \ketbra{a}{a} \otimes \tilde{\rho}_E^a,
\end{split}
\end{equation}
where
\begin{equation}
    \tilde{\rho}_E^a = (1-q) \ketbra{\psi_a}{\psi_a} + q \ketbra{\psi_{a \oplus 1}}{\psi_{a \oplus 1}}.
\end{equation}
We first calculate the eigenvalues of $\tilde{\rho}_E^0$. Since $\tilde{\rho}_E^0$ is
\begin{equation}
    \tilde{\rho}_E^0 = \left[ \begin{array}{cc}
        (1-q) + qF^2 & qF\sqrt{1-F^2} \\
        qF \sqrt{1-F^2} & q (1-F^2) 
    \end{array} \right],
\end{equation}
the eigenvalues are
\begin{equation}
    \lambda = \frac{1 \pm \sqrt{1 - 4q(1-q)(1-F^2)}}{2}.
\end{equation}
Similarly, we can calculate eigenvalues of $\tilde{\rho}_E^1$ and we find they are the same. Thus,
\begin{equation}
    \begin{split}
        H(E|A,x^*) = h \left( \frac{1 + \sqrt{1 - 4q (1-q) (1-F^2)}}{2} \right).
    \end{split}
\end{equation}

Substituting all three terms into the right-hand side of Eq.~(\ref{eq:condHrewrite}), we have
\begin{equation}\label{eq:attack}
\begin{split}
    H(A|E,x^*) &= 1 - h\left(\frac{1+F}{2}\right)\\
    &+ h\left(\frac{1 + \sqrt{1-4q(1-q)(1-F^2)}}{2}\right ).
\end{split}
\end{equation}

Next, we calculate the asymmetric parity-CHSH value $S^\text{P-CHSH}_\alpha$ for the quantum state and the measurements defined above in the following way
\begin{equation}
    \begin{split}
        S^\text{P-CHSH}_\alpha &= \alpha \ev{A_0 B_+^1} + \ev{A_1 B_-^1 \bigotimes_{i = 2}^{N-1} B_1^i }\\
        &= 2 \alpha \cos \theta \ev{\sigma_Z \sigma_Z \boldsymbol{I}} \\
        &+ 2 \sin \theta \ev{\sigma_X \sigma_X \cdots \sigma_X},
    \end{split}
\end{equation}
where $\boldsymbol{I}$ denotes the identity on $\text{Bob}_2, \ldots, \text{Bob}_{N-1}$. The expectation values can be calculated as
\begin{align}
    \ev{\sigma_Z \sigma_Z \boldsymbol{I}} &= 1\\
    \ev{\sigma_X \sigma_X \cdots \sigma_X} &= F.
\end{align}
Hence, we can relate $S^\text{P-CHSH}_\alpha$ and $F$
\begin{equation}
    \begin{split}
        S^\text{P-CHSH}_\alpha &= 2 \alpha \cos \theta + 2 F \sin \theta.
    \end{split}
\end{equation}
We optimize $\theta$ to maximize the asymmetric parity-CHSH value, obtaining
\begin{equation}
    \begin{split}
        S^\text{P-CHSH}_\alpha = 2 \sqrt{\alpha^2 + F^2}.
    \end{split}
\end{equation}
Then, we have
\begin{equation}
    F = \sqrt{(S^\text{P-CHSH}_\alpha)^2/4-\alpha^2}.
\end{equation}
By substituting this relationship into Eq.~(\ref{eq:attack}), we can see that the conditional von Neumann entropy coincides with the lower bound for $\alpha \geq 1$. For $0 < \alpha <1$, we consider that Eve performs an eavesdropping attack where she probabilistically chooses the above attack and a classical attack which gives $H(A|E, x^*) = h(q)$ at $S^\text{P-CHSH}_\alpha = 2$. The conditional von Neumann entropy for this attack coincides with the lower bound for $0 < \alpha <1$. We conclude that the lower bound is tight.

\section{IMPERFECTION ANALYSIS}\label{section:imperfection}
Using our bound on the conditional von Neumann entropy, we analyze the effects of experimental imperfections in our DI-CKA protocol. We consider three sources of imperfections: detection efficiency $\eta$, global depolarizing noise $p^G$, and local depolarizing noise $p^L$. We consider that the $N$ parties share the following $N$-partite partially entangled GHZ state
\begin{equation}
    \ket{\text{GHZ}_N} = \frac{1}{\sqrt{1+r^2}} (\ket{00\cdots 0} + r\ket{11\cdots 1}).
\end{equation}
We model Alice's measurements by
\begin{equation}
    A_x = M_{0|x} - M_{1|x},
\end{equation}
where
\begin{align}
    \Pi (\phi) &= \cos \phi \sigma_Z + \sin \phi \sigma_X,\\
    M_{0|x} &= (\Pi (\phi^x_A) + I)/2,\quad
    M_{1|x} = I - M_{0|x},
\end{align}
and $I$ denotes the identity operator. Analogously, Bobs' measurements are
\begin{equation}
    B_{y_i}^i = N_{0|y_i} - N_{1|y_i},
\end{equation}
where
\begin{align}
    N_{0|y_i} &= (\Pi (\phi_{B^i}^{y_i}) + I)/2,\\
    N_{1|y_i} &= I - N_{0|y_i},
\end{align}
for $1 \leq i \leq N-1$. We optimize the parameter $r$ of the GHZ state, the probability of noisy preprocessing $q$, and the parameter $\alpha$ to maximize key rates. We assume the parties perform $\sigma_Z$ measurement for the key-generation rounds, i.e., $A_0 = B_2^1 = \cdots B_2^{N-1} = \sigma_Z$, and $\text{Bob}_2, \ldots, \text{Bob}_{N-1}$ perform $\sigma_X$ measurements corresponding to the other input, i.e., $B_1^2 = \cdots B_1^{N-1} = \sigma_X$. We optimize the measurement angles corresponding to $A_1, B_0^1$ and $B_1^1$ to maximize key rates.

\subsection{Detection efficiency}
\begin{figure}[tbp]
\includegraphics[keepaspectratio, scale=0.45]{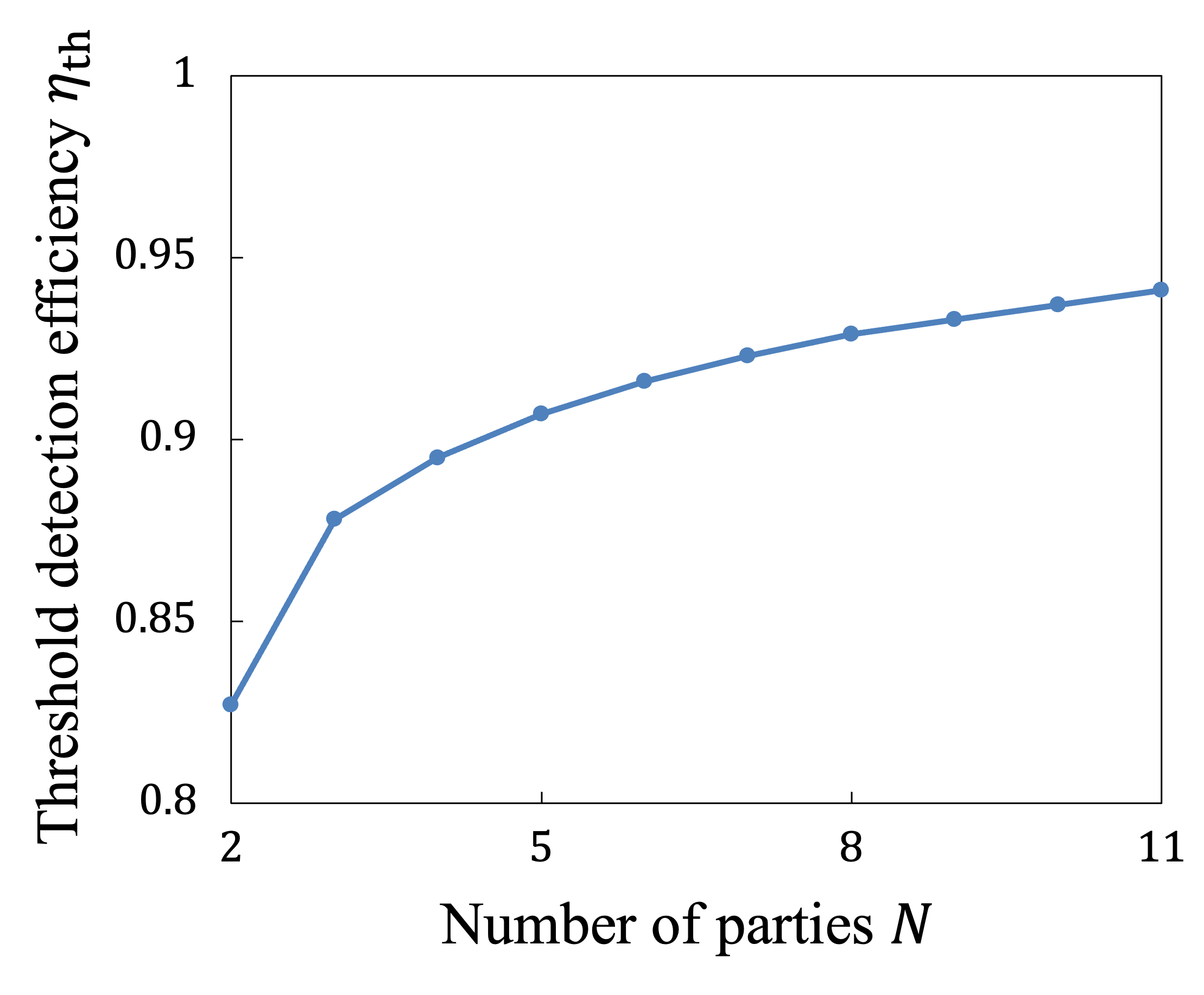}
\caption{\label{fig:DeteffvsParty}
Threshold detection efficiency $\eta_\text{th}$ against the number of parties $N$. We optimize the parameter of the distributed GHZ state $r$, the measurement angles $\phi$, and the probability of noisy preprocessing $q$ to maximize key rates. We use $\alpha = 1$.}
\end{figure}

Consider first the effect of detection efficiency $\eta$, which we model as follows
\begin{align}
    M_{0|x} = \eta (\Pi(\phi_A^x) + I)/2, &\quad
    M_{1|x} = I - M_{0|x},\\
    N_{0|y_i} = \eta (\Pi(\phi^{y_i}_{B^i}) + I)/2, &\quad
    N_{1|y_i} = I - N_{0|y_i}.
\end{align}
We analyze threshold detection efficiencies $\eta_\text{th}$, i.e., minimum detection efficiencies to obtain positive key rates. In Table~\ref{Tab:DetectionEffNoAlpha}, we show the results for $N = 2, \ldots, 6$ parties, optimizing the parameter of the GHZ state $r$ and the probability of noisy preprocessing $q$. Here, we fix $\alpha = 1$. $r_\text{fix}$ and $q_\text{fix}$ indicate that we do not optimize the parameters, i.e., we use $r = 1$ and $q = 0$, respectively. On the other hand, $r_\text{opt}$ and $q_\text{opt}$ indicate that we optimize the parameters $r$ and $q$, respectively. It can be observed that we can tolerate lower detection efficiencies by appropriately tuning the parameters $r$ and $q$. For example, when the number of parties is 3, the threshold detection efficiency $\eta_\text{th}$ is 93.4 \% with $r = 1$ and $q = 0$. However, by optimizing the parameters $r$ and $q$, it  decreases to 87.8 \%. We also observe that we require higher detection efficiencies to obtain positive key rates when the number of parties increases. In Fig.~\ref{fig:DeteffvsParty}, we plot the threshold detection efficiency $\eta_\text{th}$ against the number of parties $N$. This phenomenon may be attributed to the fact that required detection efficiency to violate the parity-CHSH inequality increases with the number of parties. However, this property of the parity-CHSH inequality contrasts with other Bell inequalities which we can violate with lower detection efficiency for a larger number of parties~\cite{Bjerrum2023}. It is an important future direction to analyze whether this property holds or not for any DI-CKA protocols.

Threshold detection efficiencies for the number of parties $N = 2, \ldots, 6$ where we optimize not only the parameters $r$ and $q$ but also the parameter $\alpha$ are shown in Table~\ref{Tab:DetectionEffAlpha}. When the number of parties is $N = 3$ and we use $r = 1$ and $q = 0$, the threshold detection efficiency decreases from 93.4 \% to 93.2 \% by optimizing $\alpha$. Although we see improvements by optimizing $\alpha$ when we optimize only the probability $q$, marginal improvements are observed when we also optimize $r$.

\begin{table}[tbp]
\centering
\caption{Threshold detection efficiency $\eta_\text{th}$ where we fix $\alpha = 1$. $r_\text{fix}$ and $q_\text{fix}$ indicate that we use $r = 1$ and $q = 0$, respectively, and $r_\text{opt}$ and $q_\text{opt}$ indicate that we optimize the parameters $r$ and $q$, respectively.}
\label{Tab:DetectionEffNoAlpha}
\begin{tabular}{c|c|c|c|c}
    $N$ & $r_\text{fix}$/$q_\text{fix}$ & $r_\text{opt}$/$q_\text{fix}$ & $r_\text{fix}$/$q_\text{opt}$ & $r_\text{opt}$/$q_\text{opt}$ \\ \hline
    2 & 92.4 \% & 88.5 \% & 91.3 \% & 82.7 \% \\
    3 & 93.4 \% & 90.6 \% & 92.4 \% & 87.8 \% \\
    4 & 94.1 \% & 91.9 \% & 93.2 \% & 89.5 \% \\
    5 & 94.6 \% & 92.7 \% & 93.8 \% & 90.7 \% \\
    6 & 95.0 \% & 93.4 \% & 94.3 \% & 91.6 \%
\end{tabular}
\end{table}

\begin{table}[bp]
\centering
\caption{Threshold detection efficiency $\eta_\text{th}$ where we optimize the parameter $\alpha$.}
\label{Tab:DetectionEffAlpha}
\begin{tabular}{c|c|c|c|c}
    $N$ & $r_\text{fix}$/$q_\text{fix}$ & $r_\text{opt}$/$q_\text{fix}$ & $r_\text{fix}$/$q_\text{opt}$ & $r_\text{opt}$/$q_\text{opt}$ \\ \hline
    2 & 92.1 \% & 88.5 \% & 91.0 \% & 82.7 \% \\
    3 & 93.2 \% & 90.6 \% & 92.2 \% & 87.8 \% \\
    4 & 93.9 \% & 91.9 \% & 93.1 \% & 89.5 \% \\
    5 & 94.5 \% & 92.7 \% & 93.7 \% & 90.7 \% \\
    6 & 95.0 \% & 93.4 \% & 94.2 \% & 91.6 \%
\end{tabular}
\end{table}

\subsection{Global depolarizing noise}
Next, we consider global depolarizing noise $p^G$, modeled as follows
\begin{equation}
    \begin{split}
        D^G(\ketbra{\text{GHZ}_N}{\text{GHZ}_N}) &= (1-p^G) \ketbra{\text{GHZ}_N}{\text{GHZ}_N} \\
        &\quad+ p^G \frac{I}{2^{N}}.
    \end{split}
\end{equation}
Again, we analyze threshold global depolarizing noise $p^G_\text{th}$ to obtain positive key rates. In Table~\ref{Tab:GlobalDep}, we show the results where we optimize the probability of noisy preprocessing $q$ and the parameter $\alpha$. Here, we fix $r = 1$ because this value is optimal for this noise. We can see that the threshold global depolarizing noise is independent of the number of parties $N$, and optimizing $q$ and $\alpha$ enhances the robustness of the protocol against this noise. When we fix the parameters $\alpha$ and $q$, the threshold global depolarizing noise is 0.142. On the other hand, the threshold becomes 0.166 by optimizing $\alpha$ and $q$.

\begin{table}[htbp]
\centering
\caption{Threshold global depolarizing noise $p^G_\text{th}$ where we fix $r=1$. $\alpha_\text{fix}$ and $q_\text{fix}$ indicate that we use $\alpha = 1$ and $q = 0$, respectively, and $\alpha_\text{opt}$ and $q_\text{opt}$ indicate that we optimize the parameters $\alpha$ and $q$, respectively.}
\label{Tab:GlobalDep}
\begin{tabular}{c|c|c|c|c}
    $N$ & $\alpha_\text{fix}$/$q_\text{fix}$ & $\alpha_\text{opt}$/$q_\text{fix}$ & $\alpha_\text{fix}$/$q_\text{opt}$ & $\alpha_\text{opt}$/$q_\text{opt}$ \\ \hline
    2 & 0.142 & 0.148 & 0.161 & 0.166 \\
    3 & 0.142 & 0.148 & 0.161 & 0.166 \\
    4 & 0.142 & 0.148 & 0.161 & 0.166
\end{tabular}
\end{table}


\subsection{Local depolarizing noise}
Finally, we consider the effect of the local depolarizing noise $p^L$. In this case, we perform the following operation for each qubit
\begin{equation}
    D^L (\rho) = (1-p^L) \rho + p^L \frac{I}{2}.
\end{equation}
We show the results in Table~\ref{Tab:LocalDep}. Similarly to the global depolarizing noise, we observe that $r = 1$ is optimal for this noise and we find that optimizing the parameter $\alpha$ and the probability $q$ can enhance the noise tolerance of the DI-CKA protocol against the local depolarizing noise. For example, we tolerate the local depolarizing noise up to 0.065 when the number of parties is $N = 3$ and we use $\alpha = 1$ and $q = 0$. On the other hand, we tolerate the local depolarizing noise up to 0.075 by optimizing $\alpha$ and $q$. Furthermore, it can be observed that the threshold $p^L_\text{th}$ decreases when the number of parties increases, indicating that the robustness deteriorates with the number of parties, as in the case of detection inefficiency.

\begin{table}[h]
\centering
\caption{Threshold local depolarizing noise $p^L_\text{th}$ where we fix $r=1$. $\alpha_\text{fix}$ and $q_\text{fix}$ indicate that we use $\alpha = 1$ and $q = 0$, respectively, and $\alpha_\text{opt}$ and $q_\text{opt}$ indicate that we optimize the parameters $\alpha$ and $q$, respectively.}
\label{Tab:LocalDep}
\begin{tabular}{c|c|c|c|c}
    $N$ & $\alpha_\text{fix}$/$q_\text{fix}$ & $\alpha_\text{opt}$/$q_\text{fix}$ & $\alpha_\text{fix}$/$q_\text{opt}$ & $\alpha_\text{opt}$/$q_\text{opt}$ \\ \hline
    2 & 0.074 & 0.077 & 0.084 & 0.087 \\
    3 & 0.065 & 0.066 & 0.074 & 0.075 \\
    4 & 0.058 & 0.059 & 0.066 & 0.067 \\
    5 & 0.053 & 0.054 & 0.060 & 0.061 \\
    6 & 0.049 & 0.050 & 0.056 & 0.057
\end{tabular}
\end{table}

\section{CONCLUSION}\label{section:conclusion}
In this paper, we derived an analytical lower bound on the conditional von Neumann entropy of Alice's measurement outcomes in a DI-CKA protocol with a novel multipartite Bell inequality called the asymmetric parity-CHSH inequality and noisy preprocessing. By considering an explicit eavesdropping attack, we showed that the derived lower bound is tight. We analyzed the effects of experimental imperfections, i.e., detection efficiency, global depolarizing noise and local depolarizing noise, and found that the asymmetry of the asymmetric parity-CHSH inequality and noisy preprocessing can significantly enhance the robustness of the DI-CKA protocol against such imperfections. As future research, deriving analytical lower bounds on the conditional von Neumann entropy for other Bell inequalities with the asymmetry and noisy preprocessing is important. For example, it is known that the Holz inequality outperforms the parity-CHSH inequality~\cite{Holz2020, Grasselli2023}. Therefore, deriving an analytical entropy bound for the Holz inequality with noisy preprocessing or developing a new multipartite Bell inequality which introduces the asymmetry into the Holz inequality may improve DI-CKA protocols further.

\begin{acknowledgments}
This work was supported by JST SPRING, Grant No. JPMJSP2123, JST Moonshot R\&D, Grant No. JPMJMS226C and Grant No. JPMJMS2061, JST CRONOS, Grant No. JPMJCS24N6, and JST ASPIRE, Grant No. JPMJAP2427. We also acknowledge support from the Danish National Research Foundation, Center for Macroscopic Quantum States (bigQ, DNRF142), the  European Union’s Horizon Europe research and innovation programme under the project ``Quantum Security Networks Partnership'' (QSNP, grant agreement no. 101114043), and from Innovation Fund Denmark (CyberQ, grant agreement no. 3200-00035B and TripleQ, grant agreement no. 3200-00012B).

\end{acknowledgments}

\appendix
\section{Relationship between winning probability of asymmetric parity-CHSH game and asymmetric parity-CHSH value}\label{appendix:relationship}
We derive the relationship between the winning probability of the asymmetric parity-CHSH game $P_\text{win}^\text{AP-CHSH}$ and the asymmetric parity-CHSH value $S_\alpha^\text{P-CHSH}$ shown in Eq.~(\ref{eq:APCHSHrelation}). First, we derive the relationship for the asymmetric CHSH inequality in Eq.~(\ref{eq:ACHSHrelation}).

The asymmetric CHSH game is a type of nonlocal game~\cite{Brunner2014}. A nonlocal game is described as follows. Let Alice and Bob be two players of this game. A referee provides a question $x \in \{0, 1  \}$ ($y \in \{ 0, 1\}$) for Alice (Bob) according to some probability distribution $\mathbf{p}(x,y)$. Then, Alice (Bob) returns an answer $a \in \{ 0, 1\}$ ($b \in \{0, 1\}$) to the referee. The referee decides whether Alice and Bob win or not according to a predicate $V(a,b|x,y)$. $V(a,b|x,y) = 1$ if and only if Alice and Bob win. Then, the winning probability of Alice and Bob is expressed as
\begin{equation}
    P_\text{win} = \sum_{x, y \in \{0, 1 \}} \mathbf{p} (x, y) \sum_{a, b \in \{ 0, 1 \}} V(a,b|x,y) P(a, b|x,y),
\end{equation}
where $P(a,b|x,y)$ denotes the probability that Alice and Bob return answers $a$ and $b$ when they receive questions $x$ and $y$, respectively. Here, the following holds for Alice's and Bob's operators
\begin{equation}
    \begin{split}
        \ev{A_x B_y} &= P(0, 0|x,y) + P(1, 1|x,y) \\
        &- (P(0,1|x,y) + P(1,0|x,y))\\
        &= 2(P(0, 0|x,y) + P(1,1|x,y)) -1.
    \end{split}
\end{equation}
Then,
\begin{align}
    P(a \oplus b = 0|x, y) &= \frac{1 + \ev{A_x B_y}}{2},\\
    P(a \oplus b = 1|x, y) &= \frac{1 - \ev{A_x B_y}}{2},
\end{align}
where $a \oplus b = a + b \, \text{mod}\, 2$. Since Alice and Bob win if and only if $a \oplus b = xy$ in the asymmetric CHSH game, the winning probability can be written as
\begin{equation}
    \begin{split}
        P_{\text{win}}^{\text{CHSH}} &= \frac{\alpha}{2(1+\alpha)} \left\{ \frac{1+\ev{A_0 B_0}}{2}+ \frac{1+\ev{A_0 B_1}}{2} \right\} \\
        &+\frac{1}{2(1+\alpha)} \left\{\frac{1+\ev{A_1 B_0}}{2} + \frac{1-\ev{A_1 B_1}}{2} \right\} \\
        &= \frac{S^\text{CHSH}_\alpha}{4(1+\alpha)} + \frac{1}{2}.
    \end{split}
\end{equation}
We can derive the relationship between the winning probability of the CHSH game $P_\text{win}^\text{CHSH}$ and the CHSH value $S^\text{CHSH}$ described in Eq.~(\ref{eq:CHSHrelation}) by setting $\alpha = 1$.

We derive the relationship for the asymmetric parity-CHSH inequality shown in Eq. (\ref{eq:APCHSHrelation}) in a similar way. The winning probability of the asymmetric parity-CHSH game is
\begin{widetext}
\begin{equation}
\begin{split}
    P_\text{win}^\text{AP-CHSH} = \sum_{x, y_1 \in \{0, 1\}} \mathbf{p} (x, y_1) \sum_{a, b_1, \bar{b} \in \{ 0,1  \}} V(a, b_1, \bar{b}|x, y_1) P(a, b_1, \bar{b}|x,y_1).
\end{split}
\end{equation}
Since the $N$ players win if and only if $a + b_1 = x (y_1 + \bar{b}) \, \text{mod} \, 2$, we explicitly write the winning probability as
\begin{equation}
\begin{split}
    P_\text{win}^\text{AP-CHSH} &= \sum_{x,y_1 \in \{0,1\}} \mathbf{p}(x,y_1) \sum_{a,b_1, \bar{b} \in \{0, 1\}}V(a,b_1,\bar{b}|x,y_1) P(a,b_1,\bar{b}|x,y_1)\\
    &= \frac{\alpha}{2(1+\alpha)} \left(P(0,0,0|0,0) + P(0,0,1|0,0) +P(1,1,0|0,0) + P(1,1,1|0,0)  \right) \\
    &+ \frac{\alpha}{2(1+\alpha)} \left(P(0,0,0|0,1) + P(0,0,1|0,1) +P(1,1,0|0,1) + P(1,1,1|0,1)  \right)\\
    &+ \frac{1}{2(1+\alpha)} \left(P(0,0,0|1,0) + P(1,1,0|1,0) +P(1,0,1|1,0) + P(0,1,1|1,0)  \right)\\
    &+ \frac{1}{2(1+\alpha)} \left(P(0,1,0|1,1) + P(1,0,0|1,1) +P(0,0,1|1,1) + P(1,1,1|1,1)  \right).
\end{split}
\end{equation}
For the expectation values of Alice's and $\text{Bob}_1$'s measurements, the following holds
\begin{equation}
\begin{split}
    \ev{A_x B_{y_1}^1} &= \sum_{a, b_1,\bar{b} \in \{ 0, 1\}} (P(a=b_1,b_1,\bar{b}|x,y_1) - P(a\neq b_1,b_1,\bar{b}|x,y_1))\\
    &= \sum_{a, b_1, \bar{b} \in \{ 0, 1\}} (2P(a=b_1, b_1, \bar{b}| x,y_1))-1.
\end{split}
\end{equation}
Therefore, the two expectation values included in Eq. (\ref{eq:AsymmetricParityCHSHinequality}) can be written as
\begin{align}
    \ev{A_0B_0^1} &= 2 (P(0,0,0|0,0) + P(0,0,1|0,0) + P(1,1,0|0,0)+P(1,1,1|0,0))-1,\\
    \ev{A_0B_1^1} &= 2 (P(0,0,0|0,1) + P(0,0,1|0,1) + P(1,1,0|0,1)+P(1,1,1|0,1))-1.
\end{align}
In addition, the other two expectation values in Eq. (\ref{eq:AsymmetricParityCHSHinequality}) can be expressed as follows
\begin{align}
    \ev{A_1B_0^1\bigotimes_{i=2}^{N-1}B^i_1} &= 2(P(0,0,0|1,0)+P(1,1,0|1,0)+P(1,0,1|1,0)+P(0,1,1|1,0))-1,\\
    \ev{A_1B_1^1\bigotimes_{i=2}^{N-1}B^i_1} &= 1-2(P(0,1,0|1,1)+P(1,0,0|1,1)+P(0,0,1|1,1)+P(1,1,1|1,1)).
\end{align}
Thus, we can derive the relationship between the winning probability of the asymmetric parity-CHSH game and the asymmetric parity-CHSH value:
\begin{equation}
    \begin{split}
        P_\text{win}^\text{AP-CHSH} &= \frac{\alpha}{2(1+\alpha)} \left\{ \frac{1 + \ev{A_0B_0^1}}{2} + \frac{1 + \ev{A_0B_1^1}}{2}\right\}\\
        &+  \frac{1}{2(1+\alpha)} \left\{\frac{1 + \ev{A_1 B_0^1 \bigotimes_{i=2}^{N-1}B^i_1}}{2} + \frac{1 - \ev{A_1 B_1^1 \bigotimes_{i=2}^{N-1}B^i_1}}{2} \right\}\\
        &= \frac{S^\text{P-CHSH}_\alpha}{4(1+\alpha)} + \frac{1}{2}.
    \end{split}
\end{equation}
The relationship between the winning probability of the parity-CHSH game and the parity-CHSH value can be obtained by setting $\alpha = 1$.
\end{widetext}

\section{Strategies saturating the maximal winning probabilities of the asymmetric CHSH game}\label{appendix:ExplicitStrategies}
Here, we consider the explicit strategies of the players to achieve the classical and quantum limits of the winning probability for the asymmetric CHSH game shown in Eqs.~(\ref{eq:ClassicalWinProbACHSH}) and (\ref{eq:QuantumWinProbACHSH}), respectively. We start with the case where Alice and Bob employ classical strategies and $\alpha \geq 1$. In this case, Alice and Bob answer bits $a = b = 0$ regardless of the questions given by the referee. Then, Alice and Bob win when the pair of questions $(x, y)$ is $(0, 0)$, $(0, 1)$, or $(1, 0)$, and the winning probability becomes
\begin{equation}
    \begin{split}
        P^\text{ACHSH}_\text{win} &=\frac{\alpha}{2(1+\alpha)} \times 2 + \frac{1}{2(1+\alpha)}\\
        &= \frac{1+2\alpha}{2(1+\alpha)}.
    \end{split}
\end{equation}

Next, when Alice and Bob still employ classical strategies but $0 < \alpha < 1$. The maximal winning probability can be achieved with the strategy where Alice answers a bit which is equal to a question, that is, $a = x$, and Bob answers a bit which is a flipped bit of his question, i.e., $b = y + 1 \, \text{mod} \, 2$. With this strategy, Alice and Bob win when the pair of questions $(x, y)$ is $(0, 1)$, $(1, 0)$, or $(1, 1)$. Therefore, the winning probability of the asymmetric CHSH game in this case is
\begin{equation}
    \begin{split}
        P^\text{ACHSH}_\text{win} &=\frac{\alpha}{2(1+\alpha)}+ \frac{1}{2(1+\alpha)} \times 2\\
        &= \frac{2+\alpha}{2(1+\alpha)}.
    \end{split}
\end{equation}

Finally, we consider the case where Alice and Bob employ quantum strategies. Let the two players share a Bell state $\ket{\psi} = (\ket{00} + \ket{11})/\sqrt{2}$. When $x = 0$, Alice performs the measurement $\{ \ket{0}, \ket{1} \}$ where the first component corresponds to $a  = 0$ and the second component corresponds to $a = 1$. On the other hand, when $x = 1$, she performs the measurement $\{ \ket{+} = (\ket{0} + \ket{1})/\sqrt{2}, \ket{-} = (\ket{0} - \ket{1})/\sqrt{2} \}$. Bob performs the measurement $\{ \ket{\theta}, \ket{\theta^\perp} \}$ when $y = 0$ where the first component corresponds to $b = 0$ and the second corresponds to $b = 1$, and he performs the measurement $\{ \ket{-\theta}, \ket{-\theta^\perp}\}$ when $y = 1$, where
\begin{align}
    \ket{\theta} &= \cos \theta \ket{0} + \sin \theta \ket{1},\\
    \ket{\theta^\perp} &= -\sin \theta \ket{0} + \cos \theta \ket{1},
\end{align}
and $\theta$ is a measurement angle which he can optimize to maximize the winning probability. In this case, the winning probability where the pair of questions is $(x, y) = (0, 0)$ can be calculated as follows
\begin{equation}
    \begin{split}
        |\bra{0} \bra{\theta} \ket{\psi}|^2 + |\bra{1}\bra{\theta^\perp} \ket{\psi}|^2 = \frac{1 + \cos 2 \theta}{2}.
    \end{split}
\end{equation}
The winning probabilities for the other pairs of questions are as follows. For $(0, 1)$, the winning probability is the same as that for the pair of questions $(0, 0)$, i.e., $(1 + \cos 2 \theta)/2$. The winning probability when $(x, y)$ is $ (1, 0)$ or $(1, 1)$ can be expressed as $(1 + \sin 2 \theta)/2$. Thus, the winning probability of the asymmetric CHSH game with this strategy is
\begin{equation}
    \begin{split}
        P^\text{ACHSH}_\text{win} &= \frac{\alpha}{2(1+\alpha)} \frac{1+\cos 2 \theta}{2} \times 2\\
        &\quad+ \frac{1}{2(1+\alpha)} \frac{1+\sin 2\theta}{2} \times 2\\
        &= \frac{\alpha \cos 2\theta + \sin 2 \theta}{2(1+\alpha)} + \frac{1}{2}\\
        &= \frac{\sqrt{1+\alpha^2} \sin (2\theta + \beta)}{2 (1+\alpha)} + \frac{1}{2},
    \end{split}
\end{equation}
where $\tan \beta = \alpha$. We optimize the angle $\theta$, obtaining
\begin{equation}
    \begin{split}
        P^\text{ACHSH}_\text{win} 
        &= \frac{\sqrt{1+\alpha^2}}{2 (1+\alpha)} + \frac{1}{2}.
    \end{split}
\end{equation}

\bibliography{DIQKD}

\end{document}